\documentclass[12pt,a4paper]{article}

\usepackage{amsmath,amsfonts,amssymb,bm}

\usepackage{epsfig}
\usepackage{graphicx} % do wczytywania rysunkow

\usepackage[hang,normalsize,bf]{subfigure} % rysuje kilka rysunkow
\begin{document}

\begin{center}

{\Huge \bf Heavy quark production \\ in photon-nucleon \\ and
  photon-photon collisions}

\vspace {0.6cm}

{\large A. Szczurek $^{1,2}$}

\vspace {0.2cm}

$^{1}$ {\em Institute of Nuclear Physics\\
PL-31-342 Cracow, Poland\\}
$^{2}$ {\em Rzesz\'ow University\\
PL-35-959 Rzesz\'ow, Poland\\}

\end{center}

\begin{abstract}
We discuss several mechanisms of heavy quark production
in (real) photon-nucleon and (real) photon - (real) photon
collisions. In particular we focuse on application of
the Saturation Model.
We discuss how to generalize the formula from virtual photon - proton
scattering and analyze threshold effects.
We discuss a possibility to measure the cross section for
$\gamma \gamma \rightarrow 2 c 2 \bar c$.
In addition to the main dipole-dipole contribution included
in a recent analysis, we propose how to calculate within
the same formalism the hadronic single-resolved contribution
to heavy quark production.
At high energies this yields a sizeable correction of about
30-40 \% for inclusive charm production and 15-20 \%
for bottom production.
We consider a subasymptotic component to the dipole-dipole
approach.
We get a good description of recently measured
$\sigma(\gamma \gamma \rightarrow c \bar c X)$.
Adding all possible contributions to
$e^+ e^- \rightarrow b \bar b X$ together removes a huge deficit
observed in earlier works but does not solve the problem totally.
Whenever possible, we compare the present
approach to the standard collinear one.
We propose how to distinguish different mechanisms by measuring
heavy quark-antiquark correlations.

\end{abstract}

\newpage

%-------------------------------------------------
\section{Introduction}
%-------------------------------------------------

The total cross section for virtual photon - proton
scattering in the region of small $x$ and intermediate $Q^2$
can be surprisingly well described by a simple parametrization
\cite{GW98} inspired by saturation effects related to
nonlinear phenomena due to gluon recombination for instance.
This is called Saturation Model (SM) in the literature.
The very good agreement with experimental data can be extended
even to the region of rather small $Q^2$ by adjusting
an effective quark mass ($m_{eff}$).
The value of the parameter found from the fit to
the photoproduction data is between the current quark mass
and the constituent quark mass.
At present there is no deep understanding of the fit value
of the parameter as we do not understand in detail the confinement
and the underlying nonperturbative effects related to large size
QCD contributions.

We shall try to extend the succesful SM parametrization to
quasi-real photon scattering. In the present analysis we shall
limit to the production of heavy quarks which is, in our opinion,
simpler and more transparent for real photons.
Here we can partially avoid the problem of the poor understanding
of the effective light quark mass, i.e. the domain of large
(transverse) size of the hadronic system emerging from the photon.

It was shown recently that the simple SM description
can be succesfully extended also to the photon-photon
scattering \cite{TKM01}.
In order to better understand the success of such a description
even more processes should be analysed on the same footing
in the same framework.

The heavy quark production is interesting also in the context
of a deficit of standard theoretical QCD predictions relative
to the experimental data as observed recently for bottom quark
production in proton-antiproton, photon-proton and
photon-photon scattering. This must be better understood in
the future because usually the charm production
in photon-photon collisions is considered
as a good tool to extract the gluon distribution in
the photon (see for instance \cite{JKR01}).

In the present paper we discuss and analyze many details
of the heavy quark production simultaneously in photon-nucleon
and photon-photon scattering. In particular we quantify
some new terms not included so far in the literature
on this subject. We discuss the range of applicability
of the SM parametrization for heavy quark production.
We put emphasis on some open unresolved problems
and propose ways of their resolving.

%---------------------------------------------------------------
\section{Heavy quark production in photon-nucleon scattering}
%---------------------------------------------------------------

In the picture of dipole scattering the cross section for
heavy quark-antiquark ($Q \bar Q$) photoproduction on
the nucleon can be written in general as
\begin{equation}
\sigma_{\gamma N \rightarrow Q \bar Q}(W) =
\int d^2 \rho d z \; |\Phi_T^{Q \bar Q}(\vec{\rho},z)|^2
\sigma_{d N}(\rho,z,W) \; ,
\label{SM_gamma_nuc}
\end{equation}
where $\Phi_T$ is (transverse) quark-antiquark photon wave
function (see for instance \cite{NZ91,GQ00}) and
$\sigma_{dN}$ is the dipole-nucleon total cross section.
Because of higher-order perturbative effects as well as
nonperturbative effects the latter cannot be calculated
in a simple way.
In the following, inspired by its phenomenological success
\cite{GW98} we shall use the Saturation Model parametrization
for $\sigma_{dN}$. It phenomenologically incorporates
color transparency at $r \ll 1/Q_s$ and saturation at $r\gg 1/Q_s$.
Here $Q_s$ is the saturation scale.
Because for real photoproduction the Bjorken-$x$ is not defined
we are forced to replace $x$ by a new appropriate variable.
The most logical choice would be to use gluon longitudinal
momentum fraction $x_g$ instead of Bjorken $x$.
This would lead to only minor modification of parameters to
describe experimental data for virtual photon - nucleon scattering.
In the following we take
\begin{equation}
\sigma_{dN}(\rho,z,W) = \sigma_0
\left [
 1 - \exp\left(- \frac{\rho^2}{4 R_0^2(x_{Q})}  \right)
\right ] \; ,
\end{equation}
where
\begin{equation}
R_0(x_{Q}) = \frac{1}{Q_0}
\left(
\frac{x_{Q}}{x_0}
\right)
^{-\lambda/2}  \; .
\end{equation}
Above we have introduced the quantity
$x_Q \equiv \frac{M_{Q \bar Q}^2(z)|_{min}}{W^2}$, where
$M_{Q \bar Q}(z)|^2_{min} \equiv \frac{m_Q^2}{z(1-z)}$.
The latter definition (prescription) will become clear after
the discussion below.

In Fig.1 we show predictions of the SM
for charm photoproduction. The dotted line represents
calculations based on Eq.(\ref{SM_gamma_nuc}). The result of
this calculation exceeds considerably the fixed target experimental data
\cite{Fermilab_ccbar,WA4_ccbar,EMC_ccbar,SLAC_ccbar,PEC_ccbar,TPS_ccbar}.
One should remember, however, that the simple formula (\ref{SM_gamma_nuc})
applies at high energies only. At lower energies one should
include effects due to kinematical threshold.
In the momentum representation this can be done by requiring:
$M_{Q \bar Q} < W$, where $M_{Q \bar Q}$ is invariant mass of the
final $Q \bar Q$ system. This cannot be done strictly
in the mixed representation used to formulate the Saturation Model
as here the information about heavy quark/antiquark momenta is
not available. An approximate way to include the obvious limitation
for heavy quark production is to neglect momenta and include only
heavy quark masses in calculating $M_{Q \bar Q}$. This upper limit
still exceeds the low energy experimental data.
There are phase space limitations in the region
$x_g \rightarrow$ 1 which has been neglected so far.
Those can be estimated using naive counting rules.
Counting reaction spectators we get for our process
an extra correction factor
\begin{equation}
S_{sup} = (1-x_g)^7
\label{threshold_function_gamma_p} \; .
\end{equation}
In our case of mixed representation we are forced to use rather
$x_Q$ instead of $x_g$.
Such a procedure leads to a reasonable agreement with the
fixed target experimental data as can be seen by comparing the
solid line and the experimental data points.

The deviation of the solid line from the dotted line gives an idea
of the range of the safe applicability of the Saturation Model for
the production of the charm quarks/antiquarks.
The cross section for $W >$ 20 GeV is practically independent
of the approximate treatment of the threshold effects. The only data
in this region come from HERA. Here the Saturation Model seems
to slightly underestimate the H1 collaboration data \cite{H1_ccbar}.

\begin{figure}

\begin{center}
\includegraphics[width=10cm]{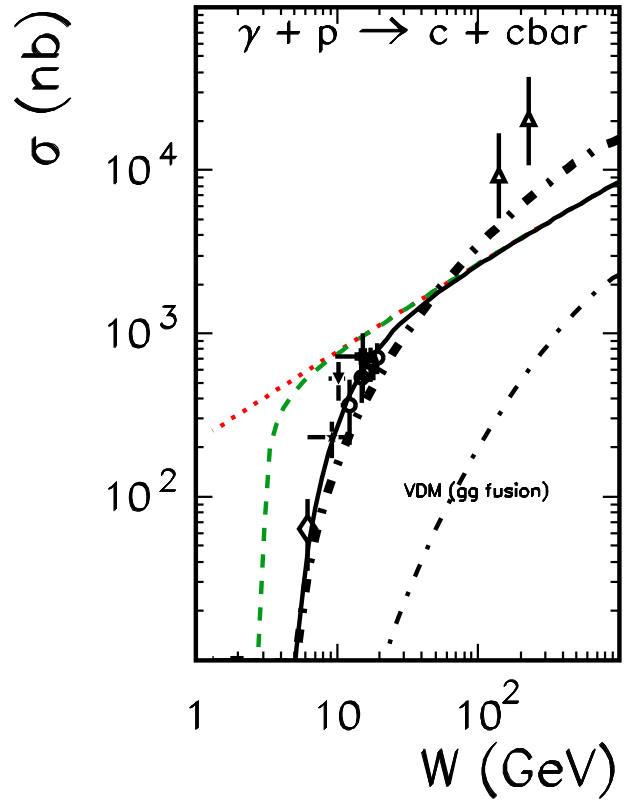}

\caption{\it The cross section for $\gamma + p \rightarrow c \bar c X$
as a function of center-of-mass photon-proton energy.
The dotted line is the result obtained with the Saturation Model,
the dashed line includes kinematical threshold and
the solid line includes in addition a suppression by the factor
$(1-x_c)^7$. The thick dashed-dotted line was obtained in the
collinear approximation and the thin dash-dotted line represents
the LO VDM contribution.
}
\end{center}

\label{fig_gp_ccbar}
\end{figure}

For comparison in Fig. 1 we show the result of
similar calculations in the collinear approach (thick dash-dotted line).
In this approach the cross section reads as
\begin{equation}
\sigma_{\gamma N \rightarrow Q \bar Q}(W) =
\int dx_N \; g_N(x_N,\mu_F^2) \;
\sigma_{\gamma g \rightarrow Q \bar Q}
(\hat W) \; ,
\label{gp_QQbar_coll}
\end{equation}
where $\hat W$ is energy in the $\gamma g$ system.
The gluon distribution in the nucleon is taken from
Ref.\cite{GRV_nucleon}. In the calculation shown in
Fig.\ref{fig_gp_ccbar} we have taken
$\mu_F^2 = \mu_R^2 = m_Q^2 + p_{\perp}^2$
with $p_{\perp}$ being the heavy quark/antiquark transverse momentum.
The traditional collinear approach gives steeper energy
dependence in the energy region considered than the SM prodictions.
In order to test both approaches more, experimental data for different
energies $W >$ 20 GeV are needed.
%The only data in this region come from Ref.\cite{H1_ccbar}.

The calculation above is not complete.
For real photons a soft vector dominance contribution due to
photon fluctuation into vector mesons should
be included on the top of the dipole contribution considered
up to now.
%\footnote{This is not fully consistent but a possible double
%counting should be negligible in practice.}.
The VDM component seems unavoidable in order to understand
the behaviour of $F_2^p - F_2^n$ at small photon virtualities
\cite{SU00}. Furthermore it seems rather a natural explanation
of the strong virtuality dependence of the structure function of
the virtual photon as measured recently \cite{virtual_photon}.

In the present calculation we include only dominant
gluon-gluon fusion component. In this approximation
\begin{equation}
\sigma_{\gamma N \rightarrow Q \bar Q}^{VDM}(W) =
\sum_V \frac{4 \pi}{f_V^2} \; \int dx_V dx_N \;
g_V(x_V,\mu^2_F) \; g_N(x_N,\mu^2_F) \; \sigma_{gg \rightarrow Q \bar
  Q}(\hat W) \; .
\label{gp_VDM}
\end{equation}
Here the $f_V$ constants describe the transition of the photon
into vector mesons ($\rho$, $\omega$, $\phi$) and are taken
in the on-shell approximation from the decay of vector mesons
into dilepton pairs \cite{SU00} taking into account finite
width corrections.
The gluon distributions in vector mesons are taken as that
for the pion \cite{GRV_pion}
\begin{equation}
g_V(x_V,\mu_F^2) = g_{\pi}(x_{\pi},\mu_F^2)
\label{g_V}
\end{equation}
and that in the nucleon from Ref.\cite{GRV_nucleon}.
For factorization scale
we take $\mu_F^2 = m_Q^2 + p_{\perp}^2$ with $p_{\perp}$ being
the heavy quark/antiquark transverse momentum.
The dash-dotted line in Fig.\ref{fig_gp_ccbar} shows the VDM contribution
calculated in the leading order (LO) approximation for
$\sigma_{gg \rightarrow Q \bar Q}$ (see e.g.\cite{book}).
The so-calculated VDM contribution cannot be neglected at high
energies. Taking into account that usually the next-to-leading
order approximation leads to an enhancement by a factor $K \sim$ 2,
the VDM contribution is as important as the continuum calculated
above.

\begin{figure}

\begin{center}
\includegraphics[width=10cm]{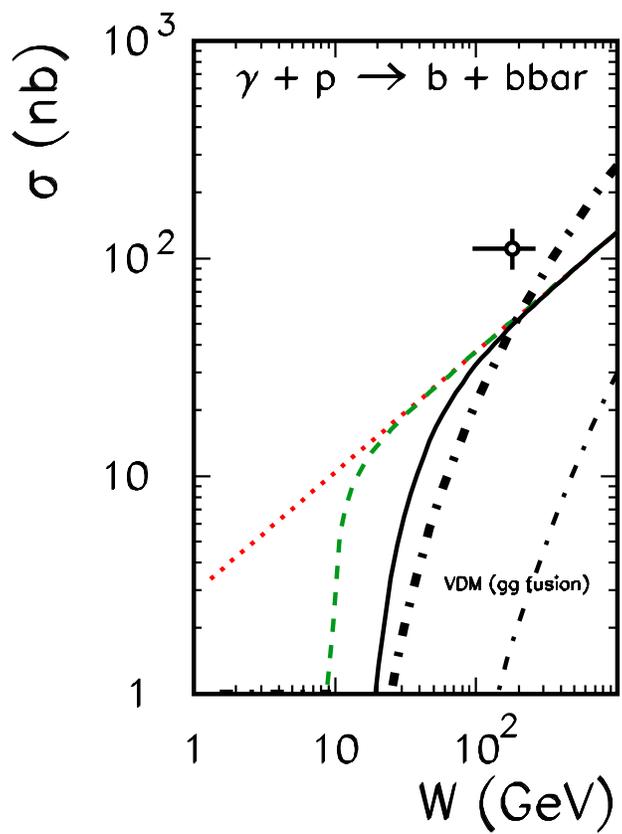}

\caption{\it  The cross section for
$\gamma + p \rightarrow b \bar b X$
as a function of the center-of-mass photon-proton energy.
The meaning of the curves is the same as in the previous figure.
The experimental data point is from \cite{H1_bbbar}.
}
\end{center}

\label{fig_gp_bbbar}
\end{figure}

The situation for bottom photoproduction seems similar.
In Fig.\ref{fig_gp_bbbar} we compare the Saturation Model predictions
with the data from the H1 collaboration \cite{H1_bbbar}.
Here the threshold effects may survive up to very high energy
$W \sim$ 50 GeV.
Again the predictions of the Saturation Model are slightly below
the H1 experimental data point. The relative magnitude
of the VDM component is similar as for the charm production.

The Saturation Model slightly underestimates high-energy data
for both charm and bottom production. The parameters of the
Saturation Model: $\sigma_0$ and $m_{eff}$ are to some
extend correlated when extracted from the fit to the experimental
data.
In principle, because in the case of heavy quark production one
is free of the uncertainty of the effective quark mass,
one could allow to modify
$\sigma_0$ to describe the HERA data on heavy quark production.
It is obvious that then one would need to increase $m_{eff}$
to describe $\sigma_{\gamma N}^{tot}$. We shall leave this
option for a separate study.

%-----------------------------------------------------------
\section{Heavy quark production in photon-photon scattering}
%-----------------------------------------------------------

%-------------------------------------------------
\subsection{Heavy quark-antiquark pair production}
%-------------------------------------------------

In the dipole-dipole approach (see for instance \cite{TKM01})
the total cross section for $\gamma \gamma \rightarrow Q \bar Q$
production can be expressed as
\begin{eqnarray}
\sigma_{\gamma \gamma \rightarrow Q \bar Q}^{dd}(W) &=&
\sum_{f_2 \ne Q} \int
                 |\Phi^{Q \bar Q}(\rho_1, z_1)|^2
                 |\Phi^{f_2 \bar f_2}(\rho_2, z_2)|^2
         \sigma_{dd}(\rho_1,\rho_2,x_{Qf}) \;
d^2 \rho_1 d z_1 d^2 \rho_2 d z_2
\nonumber \\
 &+&
\sum_{f_1 \ne Q} \int
                 |\Phi^{f_1 \bar f_1}(\rho_1, z_1)|^2
                 |\Phi^{Q \bar Q}(\rho_2, z_2)|^2
         \sigma_{dd}(\rho_1,\rho_2,x_{fQ}) \;
d^2 \rho_1 d z_1 d^2 \rho_2 d z_2 \; ,
\nonumber \\
\label{SM_QQbar}
\end{eqnarray}
where $\Phi^{q \bar q}$ are the quark-antiquark wave functions
of the photon in the mixed representation
and $\sigma_{dd}$ is dipole-dipole cross section.
While for the heavy quark-antiquark pair the photon wave function
is well defined, for light quarks one usually takes the
perturbatively calculated wave function with the quark/antiquark
mass replaced by $m_{eff}$. This parameter provides a useful
cut-off of large-size configurations in the photon wave function.

There are two problems associated
with direct use of (\ref{SM_QQbar}).
First of all, it is not completely clear how to generalize the
dipole-dipole cross section from the dipole-nucleon
cross section parametrized in \cite{GW98}.
Secondly, formula (\ref{SM_QQbar}) is correct only at sufficiently
high energy $W \gg 2 m_Q$. At lower energies one should worry about
proximity of the kinematical threshold.

In a very recent paper \cite{TKM01} a new phenomenological
parametrization for the azimuthal angle averaged dipole-dipole
cross section has been proposed:
\begin{equation}
\sigma_{dd}^{a,b}(x_{ab},\rho_1,\rho_2) =
\sigma_0^{a,b}
\left [
 1 - \exp\left(- \frac{r_{eff}^2}{4 R_0^2(x_{ab})}  \right)
\right ] \; .
%\cdot S_{thresh}(x_{ab}) \; .
\label{saturation_parametrization}
\end{equation}
Here
\begin{equation}
R_0(x_{ab}) = \frac{1}{Q_0}
\left(
\frac{x_{ab}}{x_0}
\right)
^{-\lambda/2}
\end{equation}
and the parameter $x_{ab}$ which controls the energy dependence
was defined as
\begin{equation}
x_{ab} = \frac{4 m_a^2 + 4 m_b^2}{W^2} \; .
\label{x_ab_old}
\end{equation}
In order to take into account threshold effects for the produtcion
of $q \bar q q' \bar q'$ an extra phenomenological function
has been introduced \cite{TKM01}
\begin{equation}
S_{thresh}(x_{ab}) = (1 - x_{ab})^5
\label{threshold_function}
\end{equation}
which is put to zero if $x_{ab} > 1$. This factor strongly
reduces the cross section at low energies.
Different prescriptions for $r_{eff}$ have been considered
in \cite{TKM01}, with $r_{eff}^2 = \frac{\rho_1^2 \rho_2^2}{\rho_1^2 +
  \rho_2^2}$ being the best choice phenomenologically \cite{TKM01}.

%In Ref.\cite{TKM01} it was proposed to define the
%introduced above parameters $x_{ab}$, which determine
%energy dependence, and for the case of both real photons reads as
%$x_{ab} = \frac{4 m_a^2 + 4 m_b^2}{W^2}$, i.e. is $z_1$ and $z_2$
%independent. In order to include
%threshold effects it was proposed to modify
%the cross section by a phenomenological factor $(1-x_{ab})^5$.
%However, this factor does not quarantee automatically vanishing
%of the cross section exactly below kinematical
%threshold.

The phenomenological threshold factor (\ref{threshold_function})
does not guarantee automatic vanishing of the cross section
exactly below the true kinematical threshold $W = 2 m_a + 2 m_b$.
Therefore instead of the phenomenological
factor we rather impose an extra kinematical constraint:
$M_{f \bar f} + M_{Q \bar Q} < W$ on the integration
in (\ref{SM_QQbar}).
The use of extra dynamical factor $(1-x_{ab})^5$ for heavy quark
production in photon-photon collisions is in our opinion disputable.
%In the following we shall present
%separate results with and without this factor.
Instead, in the present paper we shall estimate
the effect of damping of the cross section in the neighbourhood
of threshold due to simple kinematical limitations on the final
quark/antiquark transverse momenta. As will be discussed
in the course of this paper this leads
to a similar suppression at least numerically.

\begin{figure}

\mbox{
\epsfxsize 6.0cm
\epsffile{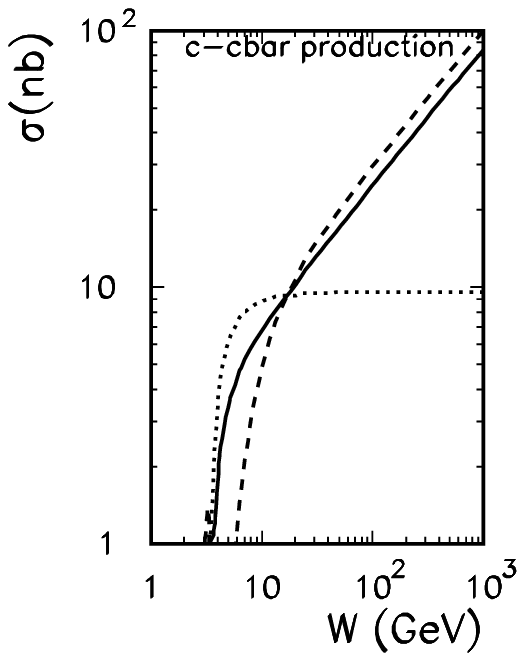}
\hspace{-1cm}
\epsfxsize 6.0cm
\epsffile{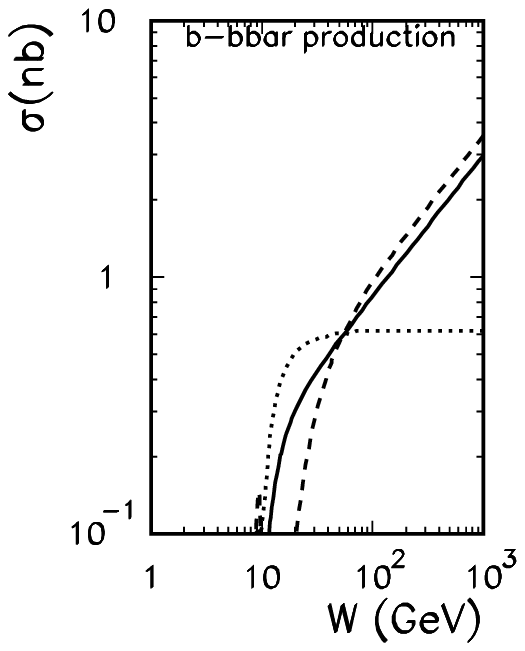}
}
\caption{\it Energy dependence of the main (see Eq.(\ref{SM_QQbar})) SM
contribution to
$\sigma(\gamma \gamma \rightarrow c \bar c X)$ (left panel)
and
$\sigma(\gamma \gamma \rightarrow b \bar b X)$ (right panel).
The dashed line corresponds to the prescription (\ref{x_ab_old})
and the solid line to the prescription (\ref{x_ab_new}).
The dotted line is the 2-gluon exchange model result
with $\alpha_s$ = 0.35.
}
\label{W_2Q}
\end{figure}

Identically as for $\gamma N$ scattering
in the mixed representation, where the transverse momenta
are not given explicitly, the quark-antiquark
invariant mass is not well defined.
We suggest therefore to use rather the effective (z-dependent!)
invariant masses $<M_{q \bar q}>$ with transverse momenta
neglected and
% where the running transverse momenta
%of quarks and antiquarks are replaced by its averaged
%value $< k_t^2 >$.
%We take $< k_t^2 >$ = 0.2$^2$, i.e.
effective quark mass as used in many other mixed representation
calculations for $\gamma N$ scattering known in the literature.
It is not completely clear how to generalize the energy
dependence of $\sigma_{d N}$ in photon-nucleon scattering
to the energy dependence in $\sigma_{dd}$ in photon-photon
scattering.
In the following we define the parameter which
controls the Saturation Model energy dependence
of the dipole-dipole cross section in a symmetric way as
%$\sigma_{dd}(\rho_1,\rho_2,x_{ab}) = (x_{ab} / x_0)^{\lambda} \cdot
% \sigma_{dd}(\rho_1,\rho_2)$ as
%
\begin{eqnarray}
x_{Qf} &=& x_{Qf}(z_1,z_2) = C \cdot
 (<M_{Q \bar Q}>(z_1) + <M_{f \bar f}>(z_2))^2 / W^2
\; ,
\nonumber \\
x_{fQ} &=& x_{fQ}(z_1,z_2) = C \cdot
 (<M_{f \bar f}>(z_1) + <M_{Q \bar Q}>(z_2))^2 / W^2
\label{x_ab_new}
\end{eqnarray}
instead of (\ref{x_ab_old}). Here we have made explicit the dependence
on $z_1$ and $z_2$. In the following we shall use C = 1 and only
in some cases compare to the results with C = 0.5.
\footnote{There is only small difference between the two
  prescriptions.}
\footnote{By construction 0 $< x_{Qf},x_{fQ} <$ 1.}
In Fig.\ref{W_2Q} we compare our prescription with those
used in \cite{TKM01} for the $c \bar c$ production (left panel)
and for the $b \bar b$ production (right panel).
In comparison to \cite{TKM01} our prescription leads
to a small reduction of the cross section far from the threshold
and a significant enhancement close to the threshold.
Some consequences of this will be discussed separately in the context
of the observed excess of $b \bar b$ production in positron-electron
collisions.

For comparison we show in Fig.\ref{W_2Q} also a result obtained
in the two-gluon exchange model. In this case
\begin{equation}
\sigma_{dd}(\vec{\rho}_1,\vec{\rho}_2) = \frac{8}{9}
\int \frac{d^2 \kappa}{\kappa^4} \alpha_s^2(\mu^2)
\cdot
(2 - e^{i \vec{\kappa} \vec{\rho}_1} - e^{-i \vec{\kappa}
  \vec{\rho}_1})
\cdot
(2 - e^{i \vec{\kappa} \vec{\rho}_2} - e^{-i \vec{\kappa}
  \vec{\rho}_2})
\; .
\label{dipole_dipole}
\end{equation}
In principle one can allow for running of $\alpha_s(\mu^2)$.
However, the choice of the scale $\mu^2$ is not completely clear.
In the present calculation a rather large constant value
$\alpha_s$ = 0.35 was taken.
If $\alpha_s(m_c^2/m_b^2)$ was used the cross section
would be negligibly small.
Thus it becomes clear that the Saturation Model leads to a huge
enhancement relative
to the two-gluon exchange model high above the threshold.
Close to the threshold both results almost coincide.
The departure of the 2g-exchange result from the constant value
below $W \approx 3 \cdot (2 m_Q)$ is due to the threshold effects.

\begin{figure}

\begin{center}

\includegraphics[width=10cm]{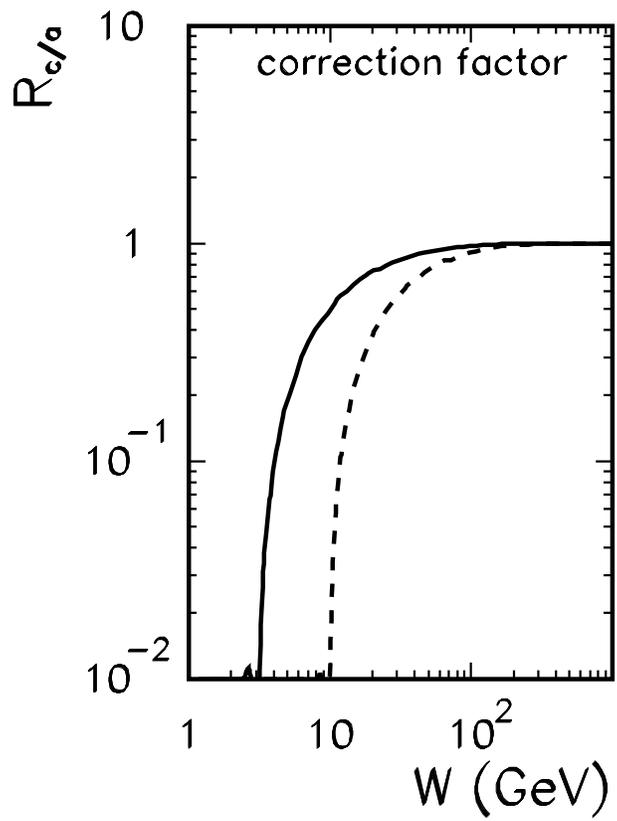}

\caption{\it The correction factor $R_{c/a}$ as a function
of W for $c \bar c$ (solid) and $b \bar b$ (dashed) production.
}
\end{center}
\label{correction_factor}
\end{figure}

In calculating the main SM component above we have considered
only obvious kinematical limitations possible to implement
in the mixed representation.
As already discussed it is not possible to include the
transverse momentum limitations in the mixed representation.
Now we shall estimate the effect due to the kinematically
limited integration over transverse momenta of final
heavy quarks/antiquarks. This effect can be taken into
account consistently only in the momentum representation.
This would require a reformulation of the whole Saturation Model
and clearly goes beyond the scope of the present analysis.
In order to gain experience we have studied first the effect of
the kinematical constraint on the transverse momenta
in the two-gluon exchange approximation in the momentum representation.
In this case we can easily both include and exclude the
kinematical limitations on transverse momenta.
Then one can approximately correct our mixed representation
calculation as
\begin{equation}
\sigma_{\gamma \gamma \rightarrow Q \bar Q}^{mixed}(W)|_{corr} =
\sigma_{\gamma \gamma \rightarrow Q \bar Q}^{mixed}(W)|_{approx}
\cdot R_{c/a}(W) \; .
\label{finite_momentum_correction}
\end{equation}
Here for brevity we have introduced the ratio:
\begin{equation}
R_{c/a}(W) =
\frac{\sigma_{\gamma \gamma \rightarrow Q \bar Q}^{2g,mom}(W)|_{corr}}
     {\sigma_{\gamma \gamma \rightarrow Q \bar
         Q}^{2g,mom}(W)|_{approx}}
\; .
\label{R_corr/approx}
\end{equation}
In Fig.\ref{correction_factor} we show the so-obtained correction factor
as a function of W for both $c \bar c$ (solid) and
$b \bar b$ (dashed) production. As can be seen from the figure
the correction is significant even far from the threshold, i.e.
leads to a significant damping of the cross section.
We shall discuss this damping in the context of the $b \bar b$ deficit
in a separate section.

%-----------------------------------------
\subsection{$2 Q 2 \bar Q$ final states}
%-----------------------------------------

The dipole-dipole approach in general, and
the Saturation Model as its particular realization,
leads to a unique prediction for the $2 Q  2 \bar Q$
(two identical heavy quarks and two identical heavy antiquarks)
production in the final state
\begin{equation}
\sigma_{\gamma \gamma \rightarrow 2 Q 2 \bar Q}^{dd}(W) = \int
|\Phi_1^{Q \bar Q}(\rho_1, z_1)|^2
|\Phi_2^{Q \bar Q}(\rho_2, z_2)|^2
\sigma_{dd}(\rho_1,\rho_2,x_{QQ})
\;
d^2 \rho_1 d z_1 d^2 \rho_2 d z_2  \; .
\end{equation}
The same prescriptions are used here as for
the $Q \bar Q$ final states in the previous section.

\begin{figure}

\mbox{
\epsfxsize 6.0cm
\epsffile{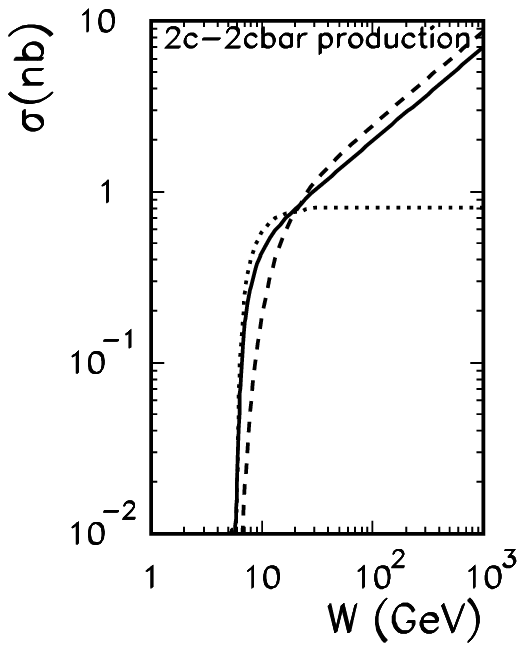}
\hspace{-1cm}
\epsfxsize 6.0cm
\epsffile{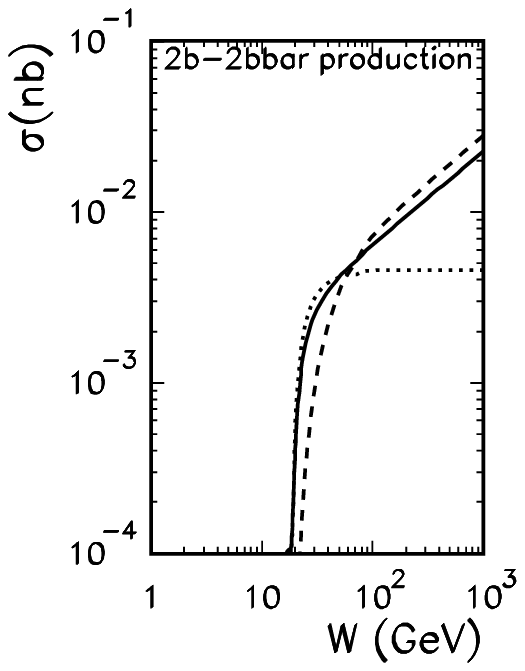}
}
\caption{\it Energy dependence of
$\sigma(\gamma \gamma \rightarrow 2c 2 \bar c)$ (left panel) and
$\sigma(\gamma \gamma \rightarrow 2b 2 \bar b)$ (right panel).
The meaning of the curves here is the same as in
the previous figure.
}
\label{W_4Q}
\end{figure}

\begin{figure}

\mbox{
\epsfxsize 6.0cm
\epsffile{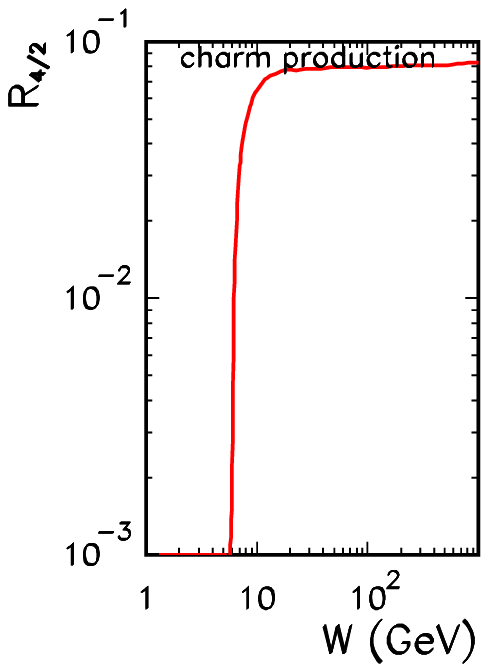}
\hspace{-1cm}
\epsfxsize 6.0cm
\epsffile{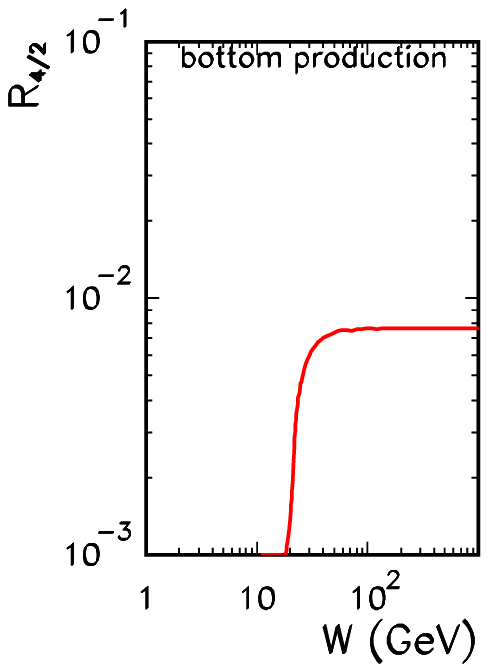}
}
\caption{\it
$R_{4/2}(W) = \frac{\sigma_{2Q 2\bar Q}(W)}{\sigma_{Q \bar Q}(W)}$
for charm (left panel) and bottom (right panel) production
as a function of photon-photon energy.
}
\label{R_4to2}
\end{figure}

In Fig.\ref{W_4Q} we compare our prescription
to that from \cite{TKM01} for four heavy quark production
for Q=c (left panel) and for Q=b (right panel).
Here there is even larger difference between the two
prescriptions than for the heavy quark-antiquark pair production.
In Fig.\ref{R_4to2} we show the ratio
%of flavour-double-tagged to flavour-single-tagged events
defined as
\begin{equation}
R_{4/2}(W) \equiv
 \frac{\sigma_{\gamma \gamma \rightarrow 2Q 2\bar Q}(W)}
      {\sigma_{\gamma \gamma \rightarrow  Q  \bar Q}(W)}
\; .
\label{R4/2}
\end{equation}
Sufficiently above the kinematical threshold $W \gg 4 m_Q$
the ratio saturates at the level of about 8 \% and 1 \% for
$2c 2\bar c$ and $2b 2\bar b$, respectively.
The fluctuations of the order of 1 \% that one can observe
in the figure are of numerical origin.

The predictions for $\gamma \gamma \rightarrow 2 Q 2 \bar Q$
shown in Fig.\ref{W_4Q} and \ref{R_4to2} are the only ones in the
literature we know. In the standard collinear approaches
the $2 Q 2 \bar Q$ final states can be produced
only in next-to-leading order calculations
or/and in the hadronization process, if the
charmed mesons, e.g. $D^*$, are measured to identify charm
quarks/antiquarks.
It would be interesting to compare in the future the present result
with the standard collinear NLO predictions.
Also from the experimental side the $2 Q 2 \bar Q$ production
in photon-photon collisions is terra incognita.
In our opinion, the $2 Q 2 \bar Q$ channel has a better chance
to be a stringent test of the dipole-dipole approaches
and the Saturation Model in particular.
It is not clear to us if the experimental verification
can be feasible with the present LEP2 statistics.
It seems, however, possible with the future photon-photon
colliders like that planned at TESLA (see for instance \cite{NIM}).

%-------------------------------------------------
\subsection{Short- versus long-distance phenomena}
%-------------------------------------------------

\begin{figure}[htbp]
  \vspace{-1cm}
  \begin{center}
    \subfigure[$\gamma_1\to c\bar{c}$]{\label{map_lt}
      \includegraphics[width=6.5cm]{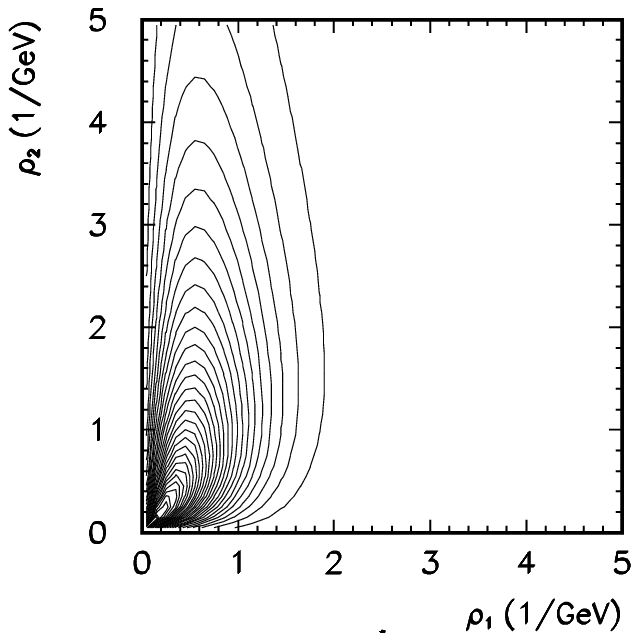}}
    \subfigure[$\gamma_2\to c\bar{c}$]{\label{map_rt}
      \includegraphics[width=6.5cm]{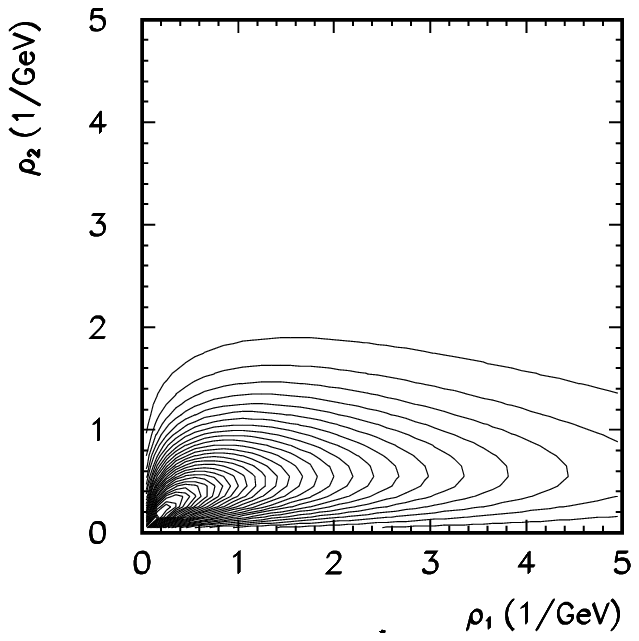}} \\
    \subfigure[light quarks only]{\label{map_lb}
      \includegraphics[width=6.5cm]{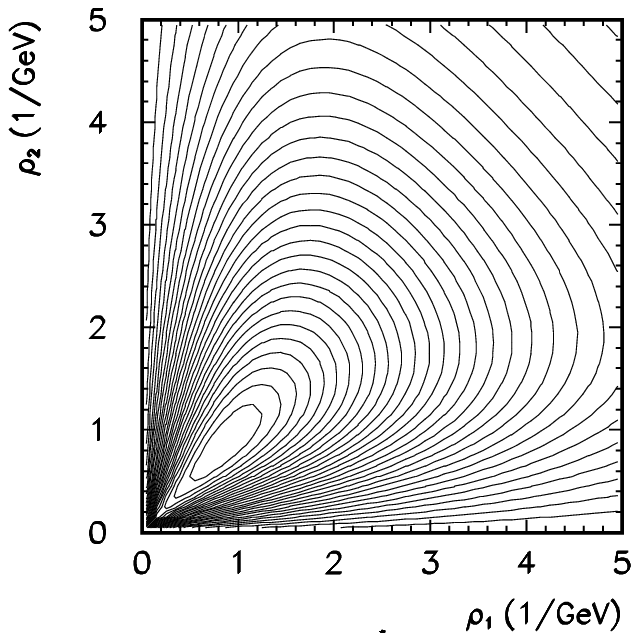}}
    \subfigure[$2c-2\bar{c}$]{\label{map_rb}
      \includegraphics[width=6.5cm]{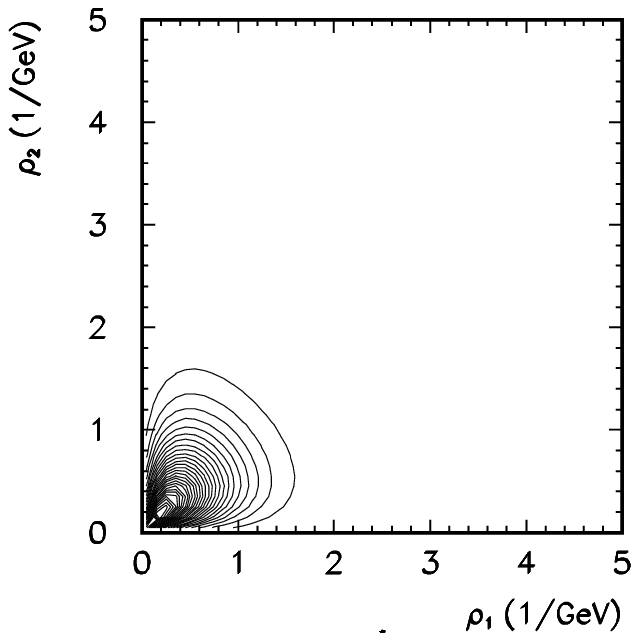}}
    \caption{\textit{A map of
      $\frac{d^2 \sigma^{\gamma \gamma \rightarrow c \bar c}(\rho_1,\rho_2)}
      {d \rho_1 d \rho_2}$
      at W = 100 GeV for the first (left-top panel) and
      the second (right-top panel) photon fluctuating
      into $c \bar c$. For comparison analogous map for light
      quark-antiquark pairs (left-bottom panel) and for
      the case when both pairs consist of charm quarks/antiquarks
      (right-bottom panel)}}
    \label{map_rho1_rho2}
  \end{center}
\end{figure}

What are typical distances probed in heavy quark production ?
Is the heavy quark production a short distance phenomenon ?
These questions can be easily answered in the mixed representation
formulation considered in the present paper.
In Fig.\ref{map_rho1_rho2} we display the integrand of
\begin{equation}
\sigma_{\gamma \gamma \rightarrow c \bar c}(W) =
\int \; I(\rho_1,\rho_2) \; d \rho_1 d \rho_2 \; .
\end{equation}
The maxima in Fig.\ref{map_rho1_rho2}
correspond to the most probable situations.
For one light ($m_u$ = $m_d$ = $m_0$, $m_s$ = $m_0$ + 0.15 GeV)
and second heavy quark-antiquark pair the map is clearly asymmetric.
One can observe a ridge parallel to
the $\rho_1$ or $\rho_2$ axis. There is not well localized
maximum. Both short and long distances are probed.

For comparison, in the bottom part of the figure,
we show similar maps when both pairs consist
of light (u,d,s) quarks/antiquarks (left-bottom) and
in the case when both pairs consist of charm quarks/antiquarks
(right-bottom). For light quarks (u,d,s) one observes a
clear maximum at $(\rho_1,\rho_2)$ = (1 GeV$^{-1}$, 1 GeV$^{-1}$)
= (0.2 fm, 0.2 fm). In this case a nonnegligible
strength extends, however, up to large distances $\rho_1$ and
$\rho_2$.
Only in the case of the production of two $c \bar c$ pairs
the cross section is dominated exclusively by short-distance
phenomena.

%In Fig.\ref{map_project} we show projections of the two-dimensional
%maps $\rho_1$ versus $\rho_2$ onto the $\rho_{c \bar c}$ axis
%(relative transverse distance between $c$ and $\bar c$)
%for different prescriptions for $\rho_{eff}$.
%The position of the maximum strongly depends on the prescription used.
%...................................

%----------------------------------------------------
\subsection{Hadronic single-resolved processes}
%----------------------------------------------------

Up to now we have calculated the contribution when
photons fluctuate into quark-antiquark pairs.
Then the final quark and antiquark carry away the whole
longitudinal momentum of the parent photon and are
predominantly emitted in the same photon hemisphere.
In the standard collinear approach one usually includes
so-called resolved contributions, when heavy quark-antiquark
pairs are created either in the photon-gluon / gluon-photon
fusion or in the gluon-gluon fusion and quark-antiquark annihilation
processes. In the first case, known as the single-resolved
process, only a small fraction of the first or the second
photon longitudinal momentum enters into the production
of heavy quark or antiquark. In the second case, known as
double-resolved process, this is true for both photons.
The arguments above demonstrate that the resolved processes
are different, at least kinematically, from those included in
the Saturation Model, or more generally in the dipole-dipole
interaction picture. This means that the resolved processes are
not included in the dipole-dipole scattering approach.
At the same time the standard collinear approach is not complete too.
Can the dipole approach be supplemented to include the hadronic resolved
processes? Let us try to answer this question by combining
a simple vector dominance model and the dipole picture.
For simplicity we shall limit to the dominant photon fluctuations
into light vector mesons: $\rho$, $\omega$ and $\phi$, only.
It is sufficient to include the light vector mesons
because the contribution of heavier mesons, as being highly
off-mass-shell, is expected to be considerably suppressed.

If the first photon fluctuates
into the vector mesons the so-defined single-resolved contribution
to the heavy quark-antiquark production can be calculated
analogically to the photon-nucleon case as
\begin{equation}
\sigma^{SR,1}_{\gamma \gamma \rightarrow Q \bar Q}(W) =
\sum_{V_1} \frac{4 \pi}{f_{V_1}^2} \int
|\Phi_2^{Q \bar Q}(\rho_2,z_2)|^2 \sigma_{V_1 d}(\rho_2,x_1) \;
d^2 \rho_2 dz_2
\; ,
\label{SR_1}
\end{equation}
where $\Phi_2^{Q \bar Q}$ is the second photon $Q \bar Q$ wave function
and $\sigma_{V_1 d}$ is vector meson - dipole total cross section.
In the spirit of the Saturation Model, we shall parametrize the latter
exactly as for the photon-nucleon case \cite{GW98}
with a simple rescaling of the normalization factor
$\sigma_0^{dV} = 2/3 \cdot \sigma_0^{dN}$.
In the present calculation $\sigma_0^{dN}$ as well as
the other parameters of SM are taken from \cite{GW98}.
Fully analogically if the second photon fluctuates into vector mesons
\begin{equation}
\sigma^{SR,2}_{\gamma \gamma \rightarrow Q \bar Q}(W) =
\sum_{V_2} \frac{4 \pi}{f_{V_2}^2} \int
|\Phi_1^{Q \bar Q}(\rho_1,z_1)|^2 \sigma_{d V_2}(\rho_1,x_2) \;
d^2 \rho_1 dz_1
\; .
\label{SR_2}
\end{equation}
This clearly doubles the first contribution (\ref{SR_1}) to the total
cross section.
Of course, if the rapidity distributions of heavy quark/antiquark
are considered, both contributions must be treated independently.
We leave the analysis of $(\eta, k_{\perp})$ distributions for
a separate study.
The integrations in (\ref{SR_1}) and (\ref{SR_2}) are not free
of kinematical constraints.
When calculating both single-resolved contributions
it should be checked additionally if
the heavy quark-antiquark invariant mass $M_{Q \bar Q}$ is
smaller than the total photon-photon energy $W$. As in the previous
cases in the mixed representation it can be done only approximately.
In analogy to $\gamma N$ scattering the most logical definition of
the parameter which controls energy dependence of
$\sigma_{V_1 d}$ or $\sigma_{d V_2}$ is:
$x_{2/1} = M_{Q \bar Q}^2(z_{2/1}) / W^2$ and
corresponds to the gluon momentum fraction in the second or
first photon (vector meson), respectively.

\begin{figure}

\mbox{
\epsfxsize 6.0cm
\epsffile{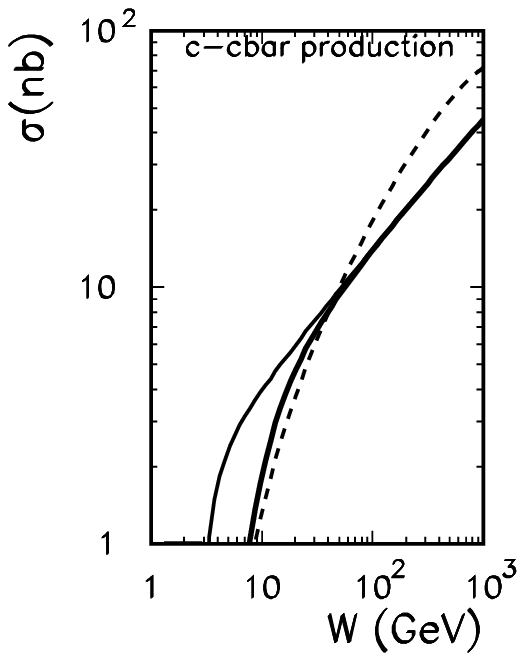}
\hspace{-1cm}
\epsfxsize 6.0cm
\epsffile{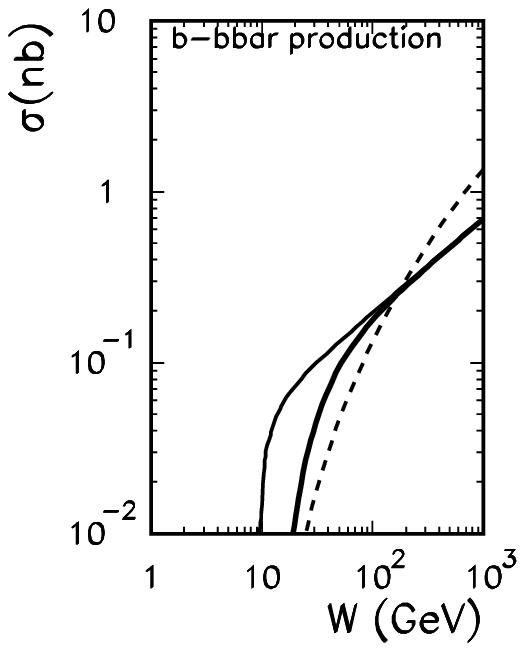}
}
\caption{\it
Hadronic single-resolved contribution to
$\gamma \gamma \rightarrow Q \bar Q X$
for $c \bar c$ (left panel) and $b \bar b$ (right panel).
The dashed line corresponds to the standard
collinear calculation as described in the text,
the solid lines correspond to the present
Saturation Model calculations without (thin solid)
and with (thick solid) the inclusion of
the suppresion factor $S_{SR}$ as obtained from
naive counting rules.
}
\label{single_resolved}
\end{figure}

In Fig.\ref{single_resolved} we compare the present result with
the result of the leading order collinear approximation.
In the latter case
\begin{eqnarray}
\sigma_{\gamma \gamma \rightarrow Q \bar Q}^{SR,coll} &=&
\sum_{V_1} \frac{4 \pi}{f_{V_1}^2} \; \int \;
g_{V_1}(x_{V_1},\mu_F^2) \;
\sigma_{g \gamma \rightarrow Q \bar Q} \; dx_{V_1}
\nonumber \\
&+&
\sum_{V_2} \frac{4 \pi}{f_{V_2}^2} \; \int \;
g_{V_2}(x_{V_2},\mu_F^2) \;
\sigma_{\gamma g \rightarrow Q \bar Q} \; dx_{V_2} \; .
\label{SR_coll}
\end{eqnarray}
Also in this calculation
the gluon distributions in the vector mesons
are taken as that for the pion \cite{GRV_pion}
$g_V(x_g,\mu_F^2) = g_{\pi}(x_g,\mu_F^2)$.
The scale $\mu_F^2 = m_c^2 + p_{\perp}^2$ is taken in practical
calculation with $p_{\perp}$ being the transverse momentum of
the final heavy quark.

There is a difference in shape of the cross section as obtained
in the SM (thin solid) and collinear (dashed) approaches.
Part of the difference may come from the fact that up to now
we have not included phase space limitations when
$x_{1/2} \rightarrow$ 1 which may be important at low energies.
In this case a naive application of counting rules would lead to
the following suppression factor
\begin{equation}
S_{SR} = (1 - x_{1/2})^{2 n_{SR}-1} \; .
\label{S_x_SR}
\end{equation}
Following standard prescription by making use of the picture
that the unresolved photon probes the resolved one and
assuming that a vector meson consists of a valence quark and antiquark
we get $n_{SR} = 3$. The SM result corrected by (\ref{S_x_SR})
is shown by the thick solid line in Fig.\ref{single_resolved}.
After the correction the results obtained from the two different
approaches are numerically fairly similar.

%----------------------------------------------------------------
\subsection{Summing all contributions to inclusive cross section}
%----------------------------------------------------------------

\begin{figure}

\mbox{
\epsfxsize 6.0cm
\epsffile{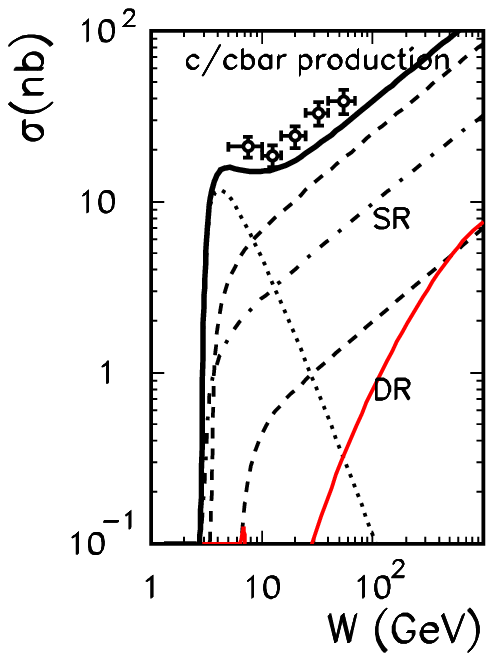}
\hspace{-1cm}
\epsfxsize 6.0cm
\epsffile{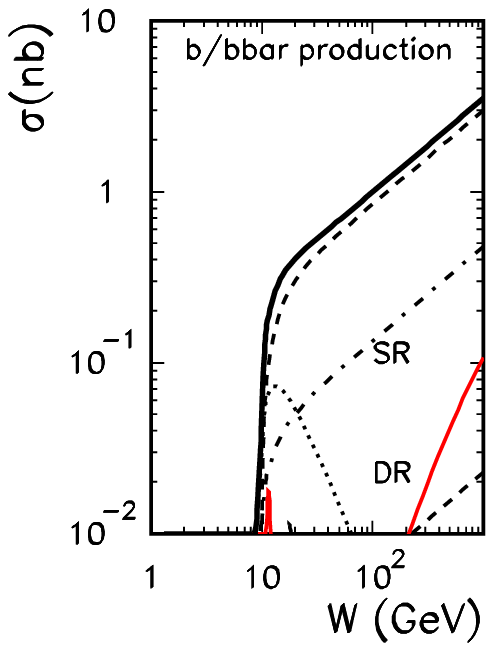}
}
\caption{\it
Different contributions to the inclusive charm (left panel)
and bottom (right panel) production in
the Saturation Model. The long-dashed line represents
the dipole-dipole contribution as proposed in \cite{TKM01} with
the modifications as described in the text, the dash-dotted line
the single-resolved contribution calculated as described in
the text and the lower dashed line the $2Q  2\bar Q$
contribution. The dotted line corresponds to the direct
contribution calculated according to \cite{Budnev}.
The double-resolved contribution is shown by the gray solid
line.
The experimental data for inclusive $c/\bar c$ production are from
Ref.\cite{L3_gg_ccbar}.
}
\label{sigma_gg_QQbar}
\end{figure}

In Fig.\ref{sigma_gg_QQbar} we show different contributions to the
inclusive $c / \bar c$ (left panel) and $b / \bar b$ (right panel)
production in photon-photon scattering.
The thick solid line represents the sum of all contributions.

Let us start from the discussion of the inclusive charm production.
The experimental data of the L3 collaboration \cite{L3_gg_ccbar}
are shown for comparison. The modifications discussed above
lead to a small damping of the cross section in comparison
to \cite{TKM01}. The corresponding result (long-dashed line)
stays below the recent experimental data of the L3 collaboration
\cite{L3_gg_ccbar}. The considered in the present paper
hadronic single-resolved contribution constitutes about 30 - 40 \%
of the main Saturation Model contribution.
At high energies the cross section for the $2c 2\bar c$
component is about 8 \% of that for the single $c \bar c$ pair
component (see Fig.\ref{W_4Q}). In inclusive
cross section its contribution should be doubled because
each of the heavy quarks/antiquarks can be potentially identified
experimentally. In principle the events with $2c 2 \bar c$ can be
subtracted both when flavour tagging is applied or charmed mesons
are measured to identify $c$ or $\bar c$.
In practice, because the efficiency of the flavour tagging is
very small and only a small fraction of $2c 2\bar c$ can be removed,
the measured inclusive cross section is two times bigger than
the integrated cross section to the $2c 2\bar c$ final state.

At higher photon-photon energies the direct contribution
is practically negligible. This is in contrast to the energy dependence
in the positron-electron collisions. Here even at large energies
the direct contribution constitutes almost half of the corresponding
cross section. The reason is that even at high $e^+ e^-$ energies the
contributions of small photon-photon energies are dominant
which can be easily understood in the equivalent photon approximation
to be discussed in the next section.
The hadronic double-resolved contribution, when each of the two
photons fluctuates into a vector meson, calculated as
\begin{equation}
\sigma_{\gamma \gamma \rightarrow Q \bar Q}^{DR,coll} =
\sum_{V_1,V_2} \frac{4 \pi}{f_{V_1}^2} \frac{4 \pi}{f_{V_2}^2}
\int g_{V_1}(x_{V_1},\mu_F^2) \; g_{V_2}(x_{V_2},\mu_F^2)
\; \sigma_{gg \rightarrow Q \bar Q}(\hat W) \;
 dx_{V_1} dx_{V_2}
\label{DR_coll}
\end{equation}
and shown by the thin solid line in the figure
becomes important only at very high energies relevant for TESLA.
Here we have consistently taken $g_V(x_V,\mu_F^2) =
g_{\pi}(x_V,\mu_F^2)$ and $\mu_F^2 = m_Q^2 + p_{\perp}^2$.

The situation for bottom production (see right panel) is somewhat
different.
Here the main SM component is dominant. Due to smaller charge
of the bottom quark than that for the charm quark the direct component
is effectively reduced with respect to the dominant SM component
by the corresponding ratio of quark/antiquark
charges: $(1/9)^2 : (4/9)^2 = 1/16$.
The same is true for the $2 b 2 \bar b$ component. Here in addition
there are threshold effects which play a role up to relatively
high energy. Also the single-resolved component is relatively
smaller which has no simple explanation.

The role of different mechanisms for charm and bottom production
is also summarized numerically in Table 1, where we have collected
corresponding cross sections for a few selected values of
photon-photon energies.

%----------------------------------------------------
\subsection{$e^+ e^- \rightarrow e^+ e^- b \bar b X$}
%----------------------------------------------------

Up to now no attempt was done to unfold experimentally
the cross section for
$\gamma \gamma \rightarrow b \bar b X$.
Only cross section for the $e^+ e^- \rightarrow b \bar b X$ reaction
was obtained recently by the L3 and OPAL collaborations
at LEP2 \cite{L3_gg_ccbar,Csilling}.
The measured, both positron and electron antitagged, cross sections
can not be described as a sum of direct and single-resolved
contributions, even if next-to-leading order corrections
are included \cite{Csilling}.
The measured cross section exceeds the theoretical predictions
by a large factor.
This is a new situation in comparison to the charm production
where the deficit is much smaller. We shall consider the
case of bottom production in more detail below.

The cross section for the $e^+ e^- \rightarrow b \bar b \; X$ reaction
when both positron and electron are antitagged
can be easily estimated in the equivalent photon approximation (EPA)
as
\begin{eqnarray}
\sigma(e^+ e^- \rightarrow b \bar b X; W_{ee}) =
\int d x_A d x_B \; f_A(E_b,\theta_{max},x_A)
                    f_B(E_b,\theta_{max},x_B)
\nonumber \\
\sigma(\gamma \gamma \rightarrow b \bar b X; W) \; ,
\label{EPA}
\end{eqnarray}
where $f_A$ and $f_B$ are virtuality-integrated flux factors
of photons in the positron and electron, respectively, and
$\theta_{max}$ is the maximal angle of the
positron/electron not to be identified by the experimental aparatus.
In the present analysis we calculate the integrated flux factors
$f_A$ and $f_B$ in a simple logarithmic approximation.
The photon-photon energy can be calculated in terms of
photon longitudinal momentum fractions $x_A$ and $x_B$ in positron
and electron, respectively, as $W = \sqrt{x_A x_B s_{ee}}$.
It is instructive to visualize how different regions of
$W_{\gamma \gamma}$ contribute to
$\sigma(e^+ e^- \rightarrow b \bar b X; W_{ee})$.
For this purpose it is useful to transform variables
from $x_A, x_B$ to $x_F \equiv x_A - x_B$ and $W_{\gamma \gamma} = W$.
Then
\begin{eqnarray}
\sigma(e^+ e^- \rightarrow b \bar b X; W_{ee}) =
\int d x_F d W \; {\cal J} \; f_A(E_b,\theta_{max},x_A)
                              f_B(E_b,\theta_{max},x_B)
\nonumber \\
\sigma(\gamma \gamma \rightarrow b \bar b X; W) \; ,
\label{EPA_W}
\end{eqnarray}
where the jacobian ${\cal J}$ is a simple function of kinematical
variables.

\begin{figure}

\mbox{
\epsfysize 5.0in
\epsffile{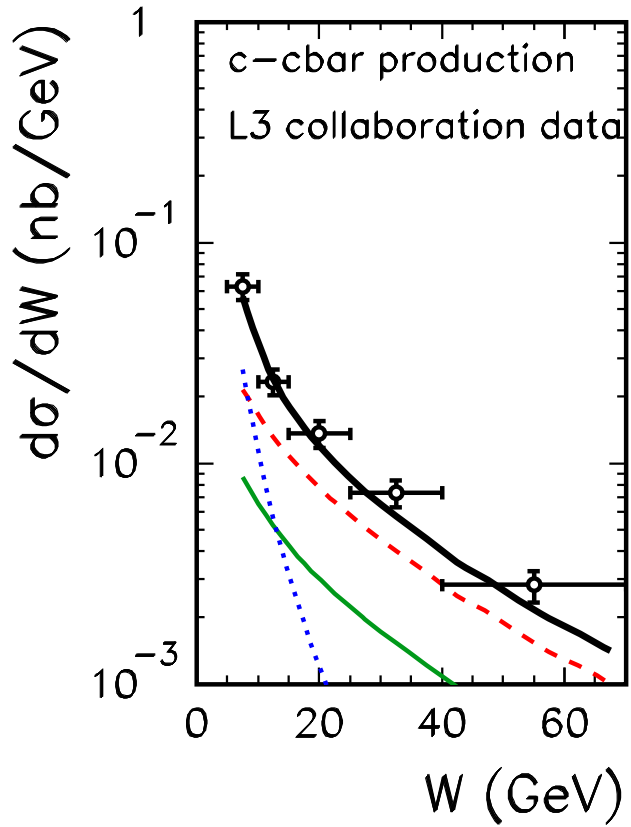}
}
\caption{\it
$ d \sigma / d W_{\gamma \gamma}$ for $e^+ e^- \rightarrow c \bar c X$.
The main SM component is shown by the dashed line,
the single-resolved component of SM by the gray solid line and
the direct component by the dotted line. The thick solid line is
a sum of all components.
The experimental data are from \cite{L3_gg_ccbar}.
}
\label{dsigma_dW_ccbar}
\end{figure}

Before we go to the $b \bar b$ production let us look at
the corresponding $c \bar c$ production.
In Fig.\ref{dsigma_dW_ccbar} we compare
the $x_F$-integrated cross section
$d \sigma / dW (e^+ e^- \rightarrow c \bar c X)$
with recent experimental data of the L3 collaboration
\cite{L3_gg_ccbar}. Different mechanisms are shown separately.
The agreement with the L3 collaboration experimental data here
is even better than for the unfolded
$\gamma \gamma \rightarrow c \bar c$ data shown
in Fig.\ref{sigma_gg_QQbar}.
This is probably due to our simplified treatment of the photon
flux factors.
The good agreement of the sum of all contributions with the data
gives credit to our model calculations.

\begin{figure}

\mbox{
\epsfxsize 6.0cm
\epsffile{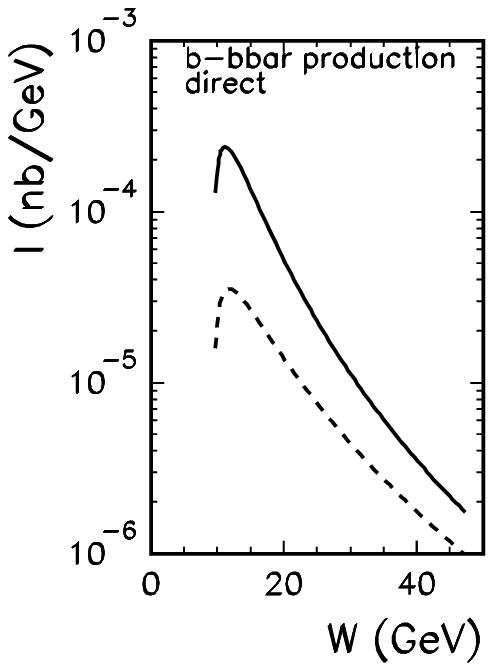}
\hspace{-1cm}
\epsfxsize 6.0cm
\epsffile{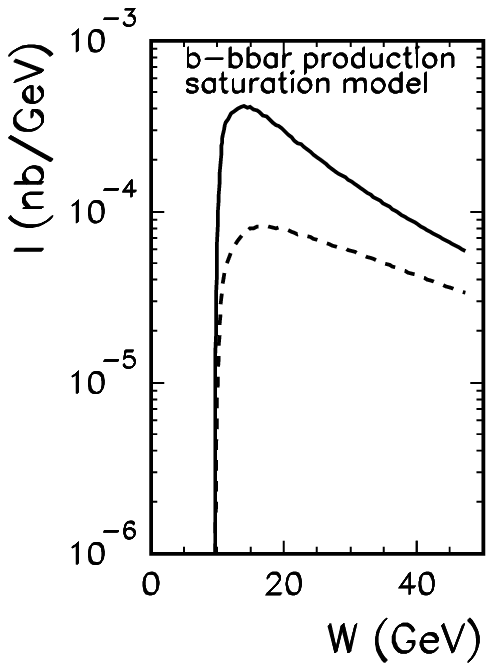}
}
\caption{\it
The dependence of the integrand of Eq.(\ref{EPA}) on
the photon-photon energy  $W_{\gamma \gamma}$ for $x_F$ = 0 (solid)
and $x_F$ = $\pm$ 0.5 (dashed) for the $b \bar b$ production for the direct
mechanism (left panel) and in the dipole-dipole scattering in
the Saturation Model (right panel) with the present
prescription for the energy dependence of the dipole-dipole cross
section. In this calculation $W_{ee}$ = 190 GeV.
}
\label{fig_wgg_dep}
\end{figure}

The integrand $I(x_F,W)$ of (\ref{EPA_W}) for $b \bar b$ is shown in
Fig.\ref{fig_wgg_dep} for the direct production (left panel) and for
the Saturation Model (right panel) including all contributions
considered in the present analysis. Quite different pattern
can be observed for both mechanisms. While for the direct production
one is sensitive mainly to low-energy photon-photon collisions,
in the Saturation Model the contributions of high-energies can not
be neglected and one has to integrate over $W_{\gamma \gamma}$
essentially up to $W_{ee}$. This difference in $W_{\gamma \gamma}$
is due to different energy dependence of
$\sigma(\gamma \gamma \rightarrow b \bar b; W_{\gamma \gamma})$
for the different mechanisms considered as has been discussed above.
But even in the latter case the integrated cross section is very
sensitive to the region of not too high-energies $W \sim$ 20 GeV,
where the not fully understood threshold effects may play essential role.

For LEP2 averaged energy
$< W_{ee} > \approx$ 190 GeV the cross section integrated
taking into account experimental cuts is
$\sigma(e^+ e^- \rightarrow b \bar b X)$ = 6.1 pb (C = 1)
or 7.4 pb (C = 1/2) for the dipole-dipole SM scattering process.
If the limitations on transverse momenta are included in addition
through the factor $R_{c/a}$ (see (\ref{R_corr/approx})), then the
corresponding cross section is reduced to 3.2 pb (C = 1)
or 3.9 pb (C = 1/2). \footnote{For comparison in the two-gluon exchange
approximation with $\alpha_s$ = 0.35 the corresponding cross
section is 10.9 pb.}
Corresponding cross section for the direct production is
$\sigma(e^+ e^- \rightarrow b \bar b X)$ = 1.2 pb.
The hadronic single-resolved contribution calculated here in
the Saturation Model as described in the present paper is very similar
in size to that calculated in the standard collinear approach
\cite{Csilling}.
As can be seen in Table 2 the $2b 2\bar b$ contribution is
practically negligible. We have completely omitted
the double-resolved contribution which is practically negligible
(see Table 1). The sum of the direct, $b \bar b$ SM,
$2 b 2 \bar b$ SM and the hadronic single resolved SM component
is 9.3-10.6 pb in the case when no transverse momenta cuts
on the main SM component are included and 6.4-7.1 pb with the cuts.
These numbers should be compared to experimentally measured
$\sigma(e^+ e^- \rightarrow b \bar b X)$ = 13.1 $\pm$ 2.0 (stat)
$\pm$ 2.4 (syst) pb \cite{L3_ee_bbbar} (L3) and preliminary
$\sigma(e^+ e^- \rightarrow b \bar b X)$ = 14.2 $\pm$ 2.5 (stat)
$\pm$ 5.0 (syst) pb \cite{L3_ee_bbbar} (OPAL).
In comparison to earlier calculations in the literature,
the theoretical deficit is much smaller.
The success of the present calculation relies on the inclusion
of a few mechanisms neglected so far, in particular
the dipole-dipole contribution which, in our opinion, is not
contained in the standard collinear approach.
%The present treatment of the threshold
%in $\gamma \gamma \rightarrow b \bar b X$ is also important in this
%context.

Up to now only $W_{\gamma \gamma}$-integrated cross section has been
determined experimentally. This, in fact, does not allow to identify
experimentally whether the problem is in low or high $W_{\gamma \gamma}$.
In order to identify better the region where the standard collinear
approach fails it would be useful to bin the experimental cross section
in the intervals of $W_{\gamma \gamma}$ making use of a possibility
to measure $W_{vis}$ which can be related to $W_{\gamma \gamma}$
via a suitable Monte Carlo program. At present, even splitting
the cross section for $e^+ e^- \rightarrow b \bar b X$ into
$\sigma(W_{\gamma \gamma} < W_0)$ and
$\sigma(W_{\gamma \gamma} > W_0)$ for $W_0 \sim$ 20 GeV
would be useful and should shed more
light on the problem of the experimental excess of $b \bar b$ relative
to the ``standard'' QCD approach.

%----------------------------------------------------
\subsection{Subasymptotic components to the dipole-dipole
approach to $\gamma \gamma \rightarrow Q \bar Q$}
%----------------------------------------------------

The Saturation Model by construction includes only
dominant asymptotic contributions relevant at
very high energies.
As discussed above the problem of heavy quark production
(especially $b \bar b$) may be a bit more complicated.
In the following we shall try to generalize the
dipole-dipole approach to include dynamics which may
be of relevance also close to threshold. As discussed
above this can be the region of the deficit of theoretical
predictions observed for the bottom production in
photon-photon collisions.

In the following as an example we shall try to estimate
the cross section for the process when the quark associated with one
photon annihilates with the same-flavour antiquark associated
with the second photon. Then the heavy quark-antiquark pair
can be produced via exchange of s-channel gluon.
In the dipole-dipole approach this effect can be
estimated(!) in terms of the familiar quark-antiquark
photon wave functions as follows
\begin{eqnarray}
\sigma_{\gamma \gamma \rightarrow Q \bar Q}^{sub} &=&
\sum_{f_1, f_2} \delta_{f_1 f_2}
\int dz_1 d^2\rho_1 dz_2 d^2\rho_2 \;
|\Psi^{f_1 \bar f_1}(z_1,\rho_1)|^2
|\Psi^{f_2 \bar f_2}(z_2,\rho_2)|^2
\nonumber \\
&\cdot&
[ \sigma_{q \bar q \rightarrow Q \bar Q}(\hat s,\vec{\rho}_1,\vec{\rho}_2) +
  \sigma_{\bar q q \rightarrow Q \bar Q}(\hat s,\vec{\rho}_1,\vec{\rho}_2) ]
\; .
\label{dd_subasymtotic}
\end{eqnarray}
The Kronecker $\delta$ reflects flavour conservation in QCD.
In the following we shall include only light flavours (u,d,s)
as ``constituents'' of both photons and use the same
treatment (suppression) of large size physics as in all cases before.
For the purpose of the estimate a sensible approximation is
to neglect light quark/antiquark transverse momenta
and write
\begin{eqnarray}
\sigma_{\gamma \gamma \rightarrow Q \bar Q}^{sub} &=&
\sum_f \int \;
q_{f/\gamma_1}^{eff}(z_1) \; \bar q_{f/\gamma_2}^{eff}(z_2)
\;
\sigma_{q \bar q \rightarrow Q \bar Q}(\hat s) \; dz_1 dz_2
\nonumber \\
&+&
\sum_f \int \;
\bar q_{f/\gamma_1}^{eff}(z_1) \;  q_{f/\gamma_2}^{eff}(z_2)
\;
\sigma_{\bar q q \rightarrow Q \bar Q}(\hat s) \; dz_1 dz_2
\; ,
\label{dd_subasymptotic_approx}
\end{eqnarray}
where $\sigma_{q \bar q \rightarrow Q \bar Q}$ or
$\sigma_{\bar q q \rightarrow Q \bar Q}$
are calculated in the Born approximation with collinear
incoming light quark and antiquark \cite{book}.
The argument of $\alpha_s(\mu_R^2)$ is set to $\mu_R^2 = \hat s$
in practical calculations.
In Eq.(\ref{dd_subasymptotic_approx}) we have introduced
for brevity
\begin{equation}
q_f^{eff}(z) \equiv \int |\Psi^{f \bar f}(z,\rho)|^2 \; d^2\rho \; .
\label{effective_quark_distribution}
\end{equation}
Because the perturbative photon wave function is singular at
$\rho$ = 0, the expression above is formally divergent. However,
there is a natural cut-off at small $\rho$'s:
$\rho_{min} \sim 1./\sqrt{\hat s}$ which makes the integral finite.
This cut-off corresponds to the upper limit on transverse momenta
in the momentum representation calculations.

\begin{figure}

\mbox{
\epsfxsize 6.0cm
\epsffile{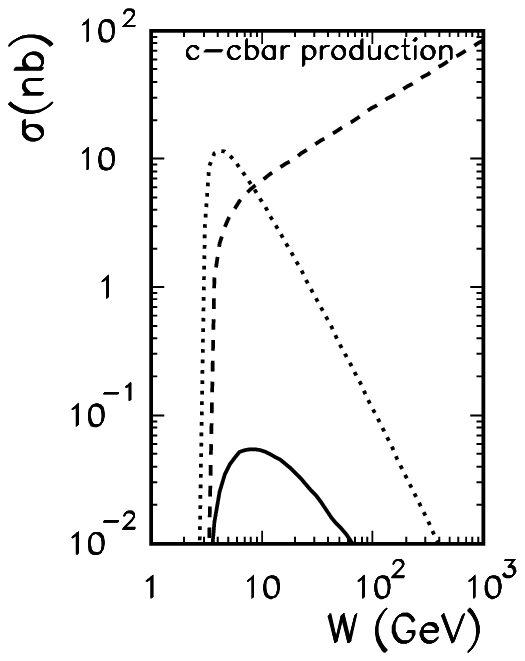}
\hspace{-1cm}
\epsfxsize 6.0cm
\epsffile{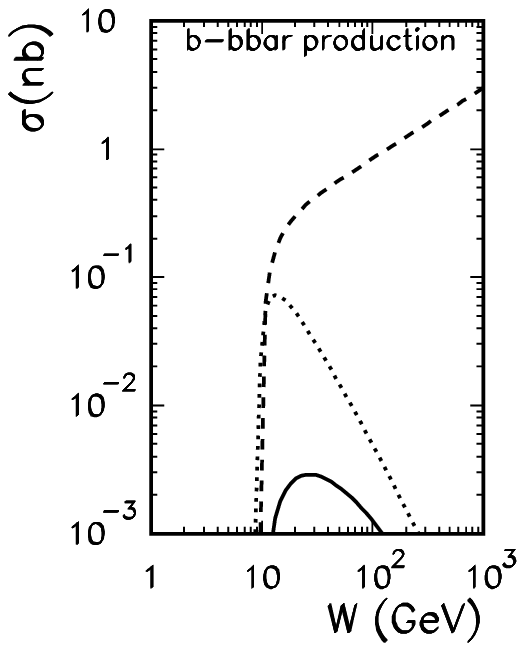}
}
\caption{\it
Subasymptotic quark-antiquark annihilation component (solid)
versus the main asymptotic component of the Saturation Model (dashed)
and the QPM component (dotted)
for charm (left panel) and bottom (right panel) production
as a function of photon-photon energy.
}
\label{dipoledipole_subasymptotic}
\end{figure}

In Fig.\ref{dipoledipole_subasymptotic} we show the cross section
for the subasymptotic contribution considered for
both $c \bar c$ and $b \bar b$ production.
The maximum of the cross section
is concentrated in the close neighbourood of corresponding
kinematical thresholds at $W \sim$ 10 GeV and $W \sim$ 30 GeV
for charm and bottom, respectively. This concentration
of the strenght at threshold justifies the introduced name
``subasymptotic'' for the process considered.
The main (asymptotic) component of SM is shown for comparison
by the dashed line and the QPM contribution by the dotted line.
The subasymptotic contribution considered shows similar energy
dependence as the QPM component.
\footnote{In fact the contribution considered can be viewed
as higher order correction to leading order QPM.}
At high energies it is rather small in comparison to
the leading asymptotic component.
Only, at threshold it constitutes a nonnegligible fraction of
the heavy quark production cross section.
When integrating $\sigma_{\gamma \gamma \rightarrow b \bar b X}^{sub}$
with virtual photon flux factors over $W_{\gamma \gamma}$ and $x_F$
we obtain $\sigma_{e^+e^- \rightarrow b \bar b X}^{sub} \approx$ 0.1 pb
at $W_{ee}$ = 190 GeV. This is rather a conservative estimate
because in the present calculations we have used leading order
formulas for $\sigma_{q \bar q \rightarrow Q \bar Q}$.
This seems tiny in comparison to the contributions considered
before (see Table 2).
We could increase slightly the cross section by a
different choice of the renormalization scale $\mu_R^2$ in
calculating $\sigma_{q \bar q \rightarrow Q \bar Q}$.
We expect that in reality
$\sigma_{e^+e^- \rightarrow b \bar X}^{sub}$ should not
exceed 0.5 pb.

%----------------------------------------
\subsection{Quark-antiquark correlations}
%----------------------------------------

So far mainly integrated cross section for heavy quark/antiquark
production was considered in the literature.
Only in a few cases inclusive distributions in transverse
momentum or rapidity (see e.g.\cite{KL96}) were presented.
No attempts were done so far to analyze the final state in more detail.
In our opinion investigating correlations between
heavy quark - heavy antiquark could be much more conclusive to
identify the production mechanisms than the integrated cross section
or even a single variable distribution.

In principle any correlation between two kinematical variables
of the final quark and antiquark would be of interest.
We suggest that the following final quark/antiquark momentum fractions:
\begin{eqnarray}
x_{Q} = \frac{\vec{p}_{Q}}
             { | \vec{p}_{Q} | } {\hat n}_{\gamma_1}
\;, \nonumber \\
x_{\bar Q} = \frac{\vec{p}_{\bar Q}}
             { | \vec{p}_{\bar Q} | }
 {\hat n}_{\gamma_1}
\; ,
\label{x_definition}
\end{eqnarray}
where
\begin{equation}
{\hat n_{\gamma_1}} =
 \frac{\vec{p}_{\gamma_1}}
 { | \vec{p}_{\gamma_1} | }
\end{equation}
would be very useful to separate the different mechanisms (approches)
analysed in the present paper. In the definition above $\vec{p}_Q$ and
$\vec{p}_{\bar Q}$ are momenta of heavy quark and antiquark,
respectively, and
$\vec{p}_{\gamma_1}$ is the momentum of the first photon,
all in the photon-photon center of mass frame.
By definition -1 $< x_{Q}, x_{\bar Q} <$ 1.
Similar quantities are being used at present when analyzing
e.g. jet production at HERA to separate out resolved and direct
processes.

\begin{figure}

\mbox{
\epsfysize 8.0in
\epsffile{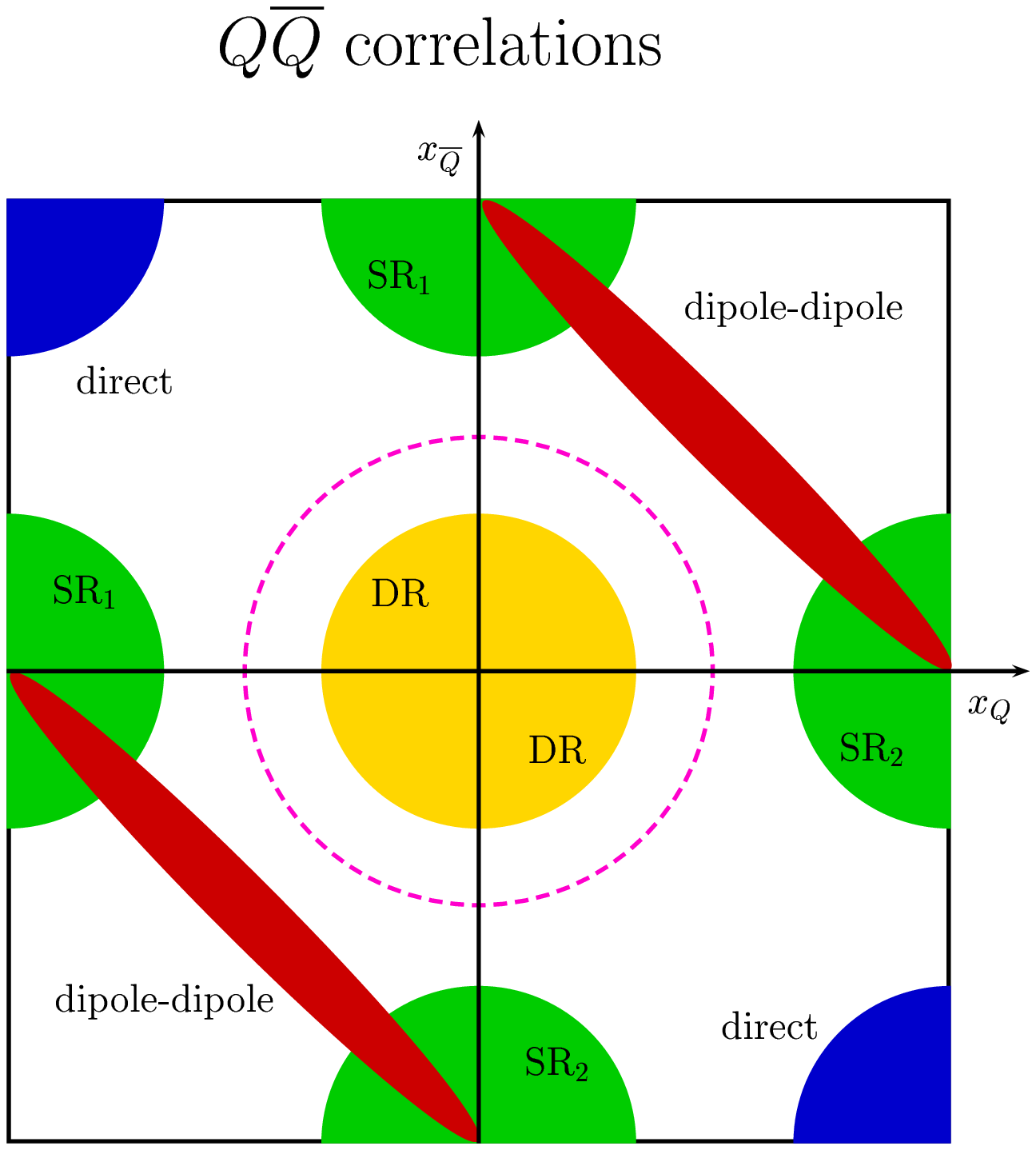}
}
\caption{\it
The expected locii in $(x_{Q},x_{\bar Q})$ space
of different mechanisms considered in the present analysis.
$SR_1$ / $SR_2$ means that the first/second photon was
transformed into vector mesons and $DR$ means that each of the
both photons was transformed into a vector meson.
The dashed circle is the locus corresponding to the pairs
emitted from the middle of the gluonic ladder (not discussed
in the text).
}
\label{correl}
\end{figure}

In Fig.\ref{correl} we present a sketch of naive expectations.
Although a precise map requires detailed calculations
for each mechanism separately, which is beyond the scope of
the present analysis, it is obvious that the separation of different
mechanisms here should be much better then for any inclusive spectra.
In the case of dipole-dipole approach (the elongated ellipses)
this would require to go to the momentum representation.
The mixed representation used in the present paper is
useful only for integrated cross sections.

Experimentally, the suggested analysis would be difficult
at LEP2 because of rather limited statistics. We hope that
such an analysis will be possible at the photon-photon option at TESLA.
At present, even localizing a few LEP2 coincidence $c \bar c$
events in the diagram $x_c$ versus $x_{\bar c}$ would be instructive.

%--------------------
\section{Conclusions}
%--------------------

There is no common consensus in the literature on detailed
understanding of the dynamics of photon-nucleon and photon-photon
collisions. In this article we have limited the discussion
to the production of heavy quarks simultaneously in
photon-nucleon and photon-photon collisions at high energies.

A special emphasis has been put on the application of
the Saturation Model which turned out recently very succesfull in
the description of experimental data for DIS at small Bjorken $x$.
We have suggested how to generalize the model to applications
with real photons. The sizeable mass of charm or bottom quarks sets
natural low energy limit on naive application of the Saturation Model.
Here a careful treatment of the kinematical threshold is required.
In the mixed representation used to formulate the Saturation Model
the effect can be included only approximately.

We have started the analysis from (real) photon-nucleon scattering,
which is very close to the domain of the
Saturation Model as formulated in \cite{GW98}.
If the kinematical threshold corrections are
included the SM gives numerically similar results as the standard
collinear approach for both charm and bottom production.
We have numerically estimated the vector dominance contribution
to the heavy quark production.

The major part of the present analysis has been devoted to
real photon - real photon collisions.
For the first time in the literature we have estimated
the cross section for the production of $2 c 2 \bar c$ final state.
Furthermore we have discussed how to include this component to
the inclusive charm production as derived in the present
experimental analyses. We have found that this component
constitutes up to 10-15 \% of the inclusive charm production
at high energies and is negligible for the bottom production.
We have shown how to generalize the Saturation Model
to the case when one of the photons fluctuates into
light vector mesons. It was found that this component yields
a significant correction of about 30-40 \% for inclusive
charm production and 15-20 \% for bottom production.
We have shown that the double resolved component, when both
photons fluctuate into light vector mesons, is nonnegligible only
at very high energies, both for the charm and bottom production.

We have shown that the production of $c \bar c$ pairs
(the same is true for $b \bar b$) is not completely
of perturbative character and involves both short- and large-size
contributions. The latter as nonperturbative are unavoidably
subjected to some modeling. Present experimental statistics do not
allow to extract cross section for the
$\gamma \gamma \rightarrow b \bar b$ reaction and therefore
it is not clear where the observed deficit reside.
It is not excluded that the apparent deficit of bottom quarks
may reside at photon-photon energies close to threshold.
This is a region where the underlying physics was never carefully
studied.
We have made a crude estimate of the subasymptotic quark-antiquark
annihilation component to $\gamma \gamma \rightarrow b \bar b$.
Although very small at high $W_{\gamma \gamma}$ its contribution
to the $e^+e^- \rightarrow b \bar b$ reaction was found to be
not completely negligible.
Adding all contributions considered in the present analysis together
removes a huge deficit observed in earlier works on
$e^+ e^- \rightarrow b \bar b X$.

The present analysis is based on the leading order impact factors.
It would be desirable in the future to perform complete
next-to-leading order calculation of heavy quark/antiquark
production in the $k_{\perp}$-factorization approach.
We expect that calculating photon impact factors
consistently up to next-to-leading order \cite{NLO_impact_factor}
may be crucial for heavy quark production.

Finally we have discussed a possibility to distniguish experimentally
the different mechanisms discussed in the present paper by measuring
heavy quark - antiquark correlations. This suggestion requires,
however, further detailed studies of the Monte Carlo type, including
experimental possibilities and limitations.

\vspace {2cm}

{\bf Acknowledments}
I am indebted to Jan Kwieci\'nski and Leszek Motyka for
the discussion of their recent work \cite{TKM01} and
Akos Csilling and Valerii Andreev for the discussion of
the recent OPAL and L3 collaboration experimental results.

\newpage

\begin{table}

\caption{Cross section for different contributions
for charm and bottom production in nb and pb, respectively.}

\begin{center}

\begin{tabular}{|c||c|c|c|c|c| }
\hline
 W (GeV) & direct & $Q \bar Q$ SM & $2 Q 2 \bar Q$ SM & SR SM & DR \\
\hline
  20  & 1.64   & 10.12  &  0.79  &  5.88   & 0.028   \\
  50  & 0.37   & 16.85  &  1.35  &  9.72   & 0.21    \\
 100  & 0.11   & 24.73  &  1.98  & 14.05   & 0.67    \\
 200  & 0.034  & 35.76  &  2.90  & 20.16   & 1.78    \\
 500  & 0.0065 & 58.67  &  4.78  & 32.06   & 5.02     \\
\hline
  20  & 53.47  & 301.9  &  0.37  & 73.06   & 0.3173(-5) \\
  50  & 14.77  & 566.3  &  4.16  &132.2   & 0.4716(-3) \\
 100  &  4.94  & 840.2  &  6.38  &196.0    & 0.3681(-2) \\
 200  &  1.56  &1228.0  &  9.38  &287.3    & 0.01719  \\
 500  &  0.32  &2047.0  & 15.56  &475.9    & 0.08679  \\
\hline
\end{tabular}

\end{center}

\end{table}

%------------------------------------------------------------

\begin{table}

\caption{Cross section in pb for $e^+ e^- \rightarrow b \bar b X$
for LEP2 averaged energy $W_{ee}$ = 190 GeV.}
\begin{center}

\begin{tabular}{ |c|c|c|c|c|c|c| }
\hline
direct & $b \bar b$ SM & $2 b 2 \bar b$ SM & SR SM & sum & L3 & OPAL \\
\hline
1.21  & 6.1-7.4 & 0.034 & 1.92  & 9.3-10.6 & 13.1 $\pm$ 2.0 $\pm$ 2.4 &
                                     14.2 $\pm$ 2.5 $\pm$ 5.0   \\
\hline

\end{tabular}

\end{center}

\end{table}

%-------------------------------------------------------------

%--------------------------------------------------------------

%-------------------------------------------------------------

%---------------------------------------------------------

%--------------------------------------------------------------

%-------------------------------------------------------------

%---------------------------------------------------------------

%-------------------------------------------------------------

%-------------------------------------------------------------

%----------------------------------------------------------------

%----------------------------------------------------------------

%-------------------------------------------------------------

%------------------------------------------------------------------

%----------------------------------------------------------------

\end{document}